\documentclass[aps,pra,showpacs,reprint]{revtex4-1}
\usepackage{hyperref}
\usepackage{amsmath}
\usepackage{braket}
\usepackage{natbib}
\usepackage{graphicx} 
\usepackage{dcolumn} 
\usepackage{amssymb}
\usepackage{color}

\begin{document}
\title{Even-odd effect in higher-order holographic production of electron vortex beams with nontrivial radial structures}
\author{G. Thirunavukkarasu}
\author{M. Mousley}
\author{M. Babiker}
\author{J. Yuan}
\affiliation{Department of Physics, University of York, Heslington, York, YO10 5DD, United Kingdom}
\date{\today}

\begin{abstract}

Structured electron beams carrying orbital angular momentum are currently of considerable interest, both from a fundamental point of view and for application in electron microscopy and spectroscopy.  Until recently, most studies have focused on the azimuthal structure of electron vortex beams with well-defined orbital angular momentum.  To unambiguously define real electron-beam states and realise them in the laboratory, the radial structure must also be specified.  Here we use a specific set of orthonormal modes of electron (vortex) beams to describe both the radial and azimuthal structures of arbitrary electron wavefronts.  The specific beam states are based on truncated Bessel beams localised within the lens aperture plane of an electron microscope.  We show that their Fourier transform set of beams can be realised at the focal planes of the probe-forming lens using a binary computer generated electron hologram.  Using astigmatic transformation optics, we demonstrate that the azimuthal indices of the diffracted beams scale with the order of the diffraction through phase amplification.  However, their radial indices remain the same as those of the encoding beams for all the odd diffraction orders or are reduced to the zeroth order for the even-order diffracted beams. This simple even-odd rule can also be explained in terms of the phase amplification of the radial profiles.  We envisage that the orthonormal cylindrical basis set of states could lead to new possibilities in phase contrast electron microscopy and spectroscopy using structured electron beams.
\end{abstract}

\maketitle

\section{Introduction}
In essence, electron microscopy can be considered as a process in which information is transferred from the sample plane to the image plane.  As the information carriers in electron microscopy are electrons, the information is encoded in the changes introduced by the sample to the amplitude and phase of the electron quantum waves.  The standard approach to the quantum theory of electron microscopy is to express the monoenergetic electron wavefunction in terms of an orthonormal basis set of states, with the information encoded in their amplitude and phase spectra.  The most commonly used quantum basis set is that of plane waves, $\Psi \left(k_x, k_y\right)$, characterized by the transverse wave-vectors $k_x$ and $k_y$. We have $k_x^2+k_y^2+k_z^2=k_0^2$, with $k_z$ the longitudinal, or axial,  wave-vector and $k_0=2\pi/\lambda$ \cite{Reimer2008}.  It also happens to be a convenient basis set as the back focal plane of the objective lens of the electron microscope then contains a map of the amplitude and phase of the plane-wave decomposition of the electron waves exiting the sample surface (plus the inevitable transfer modulation function representing the imperfection of the objective lens, a complication that needs to be considered in real applications).  However, the actual microscope in reality follows an axis-centric cylindrical design principle \cite{Hawkes1994} and the transverse extent of the electron waves is limited either by the wall of the circular electron-beam flight tube, the spatial coherence of the electron beams, or the typical round apertures used to limit the electron waves to the most spatially coherent part of the electron beam.  Thus, for the analysis of structured electron beams, an alternative quantum mechanical description of electron optics is needed which must incorporate the transversely truncated electron beam and which uses a quantum basis set involving cylindrically symmetric wave functions. 

We have recently proposed such a quantum basis set as the set of truncated Bessel functions [truncated Bessel beams (TBBs)] to describe the transverse variations of the structured electron wavefunction in the apertured regions of the beam \cite{Lloyd2017, Thirunavukkarasu2017}, such as those existing in the lens plane.  Consistent with the need to describe the transverse variations of the beam wavefunction in terms of two independent degrees of freedom, the beam modes whose transverse structure is represented by truncated Bessel beams are also characterized by two quantum numbers, namely the index $l=0, \pm1, \pm2, ...$  as the azimuthal quantum number and the index $p\ge0$ as the radial quantum number.  For non-zero $l$ , these are electron vortex beams with a topological charge $l$, carrying orbital angular momentum $l\hbar$ \cite{Lloyd2017}.

In this paper, we focus on the experimental work leading to the realisation of the set of Fourier transform (FTs) of TBBs due to higher-order diffraction from a computer generated electron hologram (CGEH).  We begin with a brief summary of the essential formalism of the TBB quantum basis set of functions and their more useful corresponding set of Fourier transforms (the FT-TBB set). Further details of the TBB and FT-TBB basis sets can be found in our recent article \cite{Thirunavukkarasu2017}.  We regard this work as laying the foundation for further developments in electron microscopy involving a cylindrically symmetric quantum basis sets.

\section{Cylindrically symmetric quantum basis}

It is well known that the wavefunction of an electron beam inside a typical electron microscope can be described adequately by Schr{\"o}dinger's equation (see for example, \cite{Reimer2008})
\begin{equation}
\mathcal{H}\ket{\Psi}=E\ket{\Psi},
\end{equation}						
where the non-relativistic Hamiltonian for an electron in free space is simply
\begin{equation}
\mathcal{H}=-\frac{\hbar^2}{2m}\bigtriangledown^2.
\end{equation}	
We note that this linear  Schr\"odinger equation is appropriate for the current generation of electron microscopies where the beam density is sufficiently small so that only a single electron needs be taken into account at any one time \cite{Lloyd2012c}.  In the case of physical scenarios where the  current density becomes sufficiently large, the nonlinear Schr\"odinger equation is required to properly describe the behaviour of the system.  The electron vortex beams we focus on here extend the study of matter vortices to the realm of practical electron microscopy in which it is now clear that such electron vortex beams bear much resemblance to optical vortices primarily as regards the singularity in phase leading to a zero amplitude on the beam axis, but differ from optical vortices in that, unlike photons, electrons bear mass and charge and have half-integer spin. Recent studies have identified vortex profiles with an amplitude singularity, rather than zero amplitude in a two-dimensional nonlinear model, where the interplay of the repulsive nonlinearity (typical for atomic Bose-Eintein Condensates) and a pull to the center by an external potential gives rise to vortices with an integrable singularity at the center (i.e., vortices with a finite total norm) \cite{Sakaguchi2011}.  In what follows we do not consider such vortex effects any further and concentrate on electron vortex beams which are governed by the linear Schr\"odinger equation.  

The  Hamiltonian of physical systems possessing cylindrical symmetry can be expressed in terms of cylindrical polar coordinates by writing, for the Laplace operator,
\begin{equation}
\bigtriangledown^2=\frac{\partial^2}{\partial \rho^2}+\frac{1}{\rho}\frac{\partial}{\partial \rho}+\frac{1}{\rho^2}\frac{\partial^2}{\partial \phi^2}+\frac{\partial^2}{\partial z^2},
\end{equation}	
where the transverse radial variable $\rho$ and the azimuthal variable $\phi$  are related to the $x$ and $y$ components of the position vector in Cartesian coordinates by $\rho=\sqrt{x^2+y^2}$ and $\phi=\arctan{(x/y)}$, respectively.
An arbitrary electron wave passing through a region defined by a circular aperture of radius $\rho_{\text{max}}$ has a wavefunction that can be written as a  superposition of the orthonormal set of functions
\begin{equation}
\Phi_l(\rho, \phi, z)=N_{l}J_l(k_{\perp}\rho)e^{il\phi}e^{ik_zz}		\,\,\, \textnormal{      for     }\rho \leq \rho_{\text{max}},
\label{Eq_TBB_wf}
\end{equation}
where $e^{ik_zz}$ is the kinetic phase factor with $k_z$ the longitudinal (axial) wavevector, which is related to the electron wave vector  $k_0$ by the relation $k_0^2=k_z^2+k_{\perp}^2$. The electron wavevector $k_0$ is related to the energy of the beam by $E=\frac{\hbar^2k_0^2}{2m}$ and $k_{\perp}=\sqrt{k_x^2+k_y^2}$ is the magnitude of the transverse (in-plane) wavevector which takes discrete (quantized) values $k_{\perp}^{pl}$ (where $p$ is the corresponding radial quantum index), depending on the boundary conditions to be detailed below.  

The transverse variations of the beam at the aperture plane are described by the azimuthal phase factor $e^{il\phi}$ and the radial function $J_l(k_{\perp}^{pl}\rho)$.  By virtue of the cylindrical symmetry, the wavefunction can only be a single-valued function if $l$ is quantized so that it only takes integer values.  This is the origin of the azimuthal index $l$.  To determine the origin of the radial index $p$, we substitute the function given in Eq.(\ref{Eq_TBB_wf}) back into the Schr\"odinger equation expressed in cylindrical coordinates [Eqs. (1)-(3)] to obtain
\begin{equation}
\frac{d^2J_l(k_\perp\rho)}{d\rho^2}+\frac{1}{\rho}\frac{dJ_l(k_\perp\rho)}{d\rho}-\frac{l^2}{\rho^2}J_l(k_\perp\rho)+(k_0^2-k_z^2)J_l(k_\perp\rho)=0.
\end{equation}
The expression on the left-hand side can be written into a more familiar form on writing $\xi=k_{\perp}\rho$ and expressing  $k_0^2-k_z^2$ as $k_{\perp}^2$.  We then have, after dividing the equation by $k_{\perp}^2$,
\begin{equation}
\frac{d^2J_l(\xi)}{d\xi^2}+\frac{1}{\xi}\frac{dJ_l(\xi)}{d\xi}+(1-\frac{l^2}{\xi^2})J_l(\xi)=0.
\end{equation}

This is the well-known Bessel differential equation whose solutions are Bessel functions of order $l$, either of the first kind or of the second kind \cite{Abramowitz1972}.  The  dependence is self-evident.  The Bessel functions of the second kind go to infinity at the origin, hence Bessel functions of the second kind cannot represent the physical wave function of an electron beam, so for our purpose the relevant solutions are the Bessel functions of the first kind which are finite everywhere.  If $\rho_\text{{max}}$ can be considered to be sufficiently large, the solutions are Bessel beams \cite{Grillo2014a, Durnin1987} characterized by the scaling parameter $k_{\perp}$ which can take any real value.   

\begin{figure}
\includegraphics[width=0.95\columnwidth]{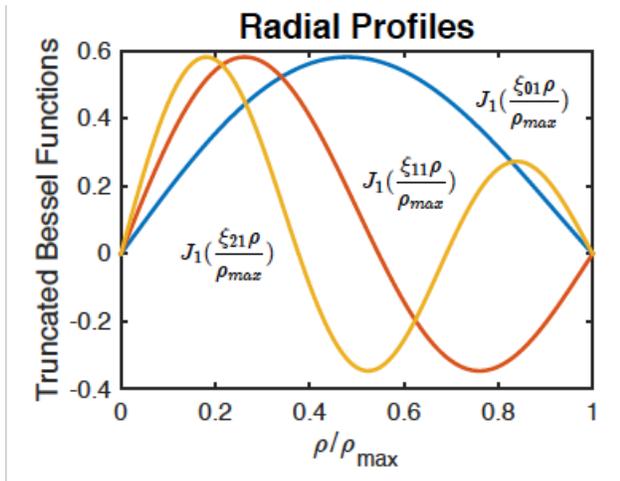}%
\caption{Examples of the radial profiles of the truncated Bessel beams (blue for $\Phi_{01}^{\textnormal{TBB}}$, orange for $\Phi_{11}^{\textnormal{TBB}}$ and yellow for $\Phi_{21}^{\textnormal{TBB}}$. Here $J_l(k_{\perp}^{pl}\rho)$ is the $p$th truncated Bessel function of the order $l$ and $\rho$ the radial variable in the aperture plane with $\rho_{\text{max}}$ the radius of the circular aperture.}
\label{FIG_RadialWaveFunctions}
\end{figure}

When $\rho_\text{{max}}$ is finite due to the presence of an aperture the solutions are TBBs, as explained earlier, with their transverse variations confined within the apertured region about the optical axis, a situation that is very common in electron microscopy because of the finite coherence of the electron sources employed, compared to the near-perfect coherence of the light sources such as that of a laser.  In that case, the scaling parameter $k_{\perp}$ takes discrete values since the Bessel functions have to conform with the boundary condition:
\begin{equation}
J_l(k_{\perp}\rho_\text{{max}})=0,
\end{equation}
which is satisfied when
\begin{equation}
k_{\perp}^{pl}\rho_\text{{max}}=\xi_{pl},
\end{equation}
with $\xi_{pl}$ the $(p+1)$th zero of $J_l(\xi)$,  the Bessel functions of order $l$.  As the Bessel functions of the first kind are oscillatory functions of their argument, very much like the sinusoidal functions, although with a gradually diminishing amplitude as $\xi$ increases, there are many such zeros for each Bessel function $J_l(\xi)$.  This leads to a set of functions describing the transverse variations of the beams which we call the truncated Bessel beams (TBB) as described in Eq. (\ref{Eq_TBB_wf}).  They are characterized by the azimuthal index $l$ and the radial index $p$.  The radial variations of lowest-order TBBs, namely, $\Phi_{01}^{\textnormal{TBB}}$(blue), $\Phi_{11}^{\textnormal{TBB}}$(orange), and $\Phi_{21}^{\textnormal{TBB}}$(yellow), are shown in Fig. \ref{FIG_RadialWaveFunctions}.

Physically, it is well known that each of the cylindrical wave functions of the form given in Eq. (\ref{Eq_TBB_wf}) has a vortex at the optical axis ($\rho=0$) if $l$ is non-zero, because of the phase indeterminacy \cite{Lloyd2017}. The functions are also eigenfunctions of the operator $\mathcal{L}_z=-i\hbar\frac{\partial}{\partial \phi}$ such that
\begin{equation}
\mathcal{L}_z \ket{\Phi_l}=l\hbar\ket{\Phi_l}.
\end{equation}
Therefore, it is clear that each member of the truncated Bessel beam set possesses an orbital angular momentum.  

As $\mathcal{L}_z$ commutes with the Hamiltonian $\mathcal{H}$, the orbital angular momentum is conserved about the beam axis, with its value proportional to azimuthal index $l$ \cite{Mousley2017}.  Consistent with this, we have shown for TBBs defined in the structured aperture plane of an electron lens that the corresponding wave functions observable at the focal plane of the electron lens are the Fourier transform of TBBs, which can be written as \cite{Thirunavukkarasu2017}
\begin{equation}
\Psi_{\text{pl}}^{\textnormal{FT-TBB}}(k_{\perp}, \varphi, k_z)=i^l\xi_{pl}J'_l(\xi_{pl})\frac{J_l(k_{\perp}\rho_{\textnormal{max}})}{(k_{\perp}^{pl})^2-k_{\perp}^2} e^{il\varphi}
\end{equation}
and are eigenfunctions of orbital angular momentum with eigenvalue $l\hbar$.

The radial index $p$ has been called the missing and ignored quantum number \cite{Plick2013, Karimi2014, Plick2015}, as it has received very little attention compared with that of the azimuthal index $l$, which current studies of electron vortex beams have largely focused on \cite{Lloyd2017}.  However, both the radial and azimuthal variations are required to fully represent any given two-dimensional transverse structure of an electron beam, particularly in a cylindrically symmetric optical systems as in the case of electron microscopy.

The study of the truncated Bessel beams using a binary computer generated hologram was discussed for the $p=0$ mode by Clark \textit{et al.} \cite{Clark2012} in the context of the production of a focused vortex beam using a lens with an aperture.  Thirunavukkarasu \textit{et al.} \cite{Thirunavukkarasu2017} presented a detailed study of the electron vortex beam generation in a similar setting, but now involving nontrivial higher-$p$-index modes.  Due to space limitations, the focus of that paper was on the first-order diffraction using a CGEH method with the results obtained equivalent to a Fourier transform of the TBB mode. 

In this paper we focus on electron vortex beams generated by high-order diffraction from the same CGEH. The effect of phase amplification present in higher-order diffraction of a computer generated hologram \cite{Endo1979} is examined in detail for a CGEH, as it is known that it can be used to generate vortex beams with high azimuthal indices \cite{McMorran2011}. The interesting question is what would their effect be on the radial indices of the vortex beams generated.%

\section{Experimental} 
\begin{figure}
\includegraphics[width=1\columnwidth]{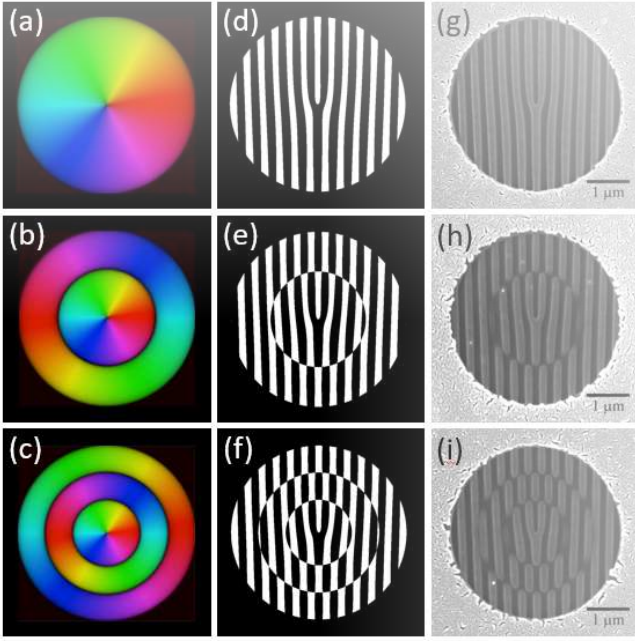}%
\caption{Phase plots of the transverse structure of the truncated Bessel beams (a) $\Phi_{01}^{\text{TBB}}$, (b) $\Phi_{11}^{\text{TBB}}$, and (c) $\Phi_{21}^{\text{TBB}}$. (d)-(f) Corresponding binary transmission functions. (g)-(i) Secondary electron microscopy images of the related amplitude masks.  The intensities plotted have been rescaled to be comparable to each other.}
\label{FIG_TransmissionMasks}
\end{figure}

The CGEH pattern is generated numerically using the standard approach \cite{Lloyd2017, Verbeeck2010} to calculate the transmission function,  defined as the amplitude function of the superposition of the beams with the desired truncated Bessel function as its transverse structure at the aperture plane $\Phi_{\text{pl}}^{\text{TBB}}(\rho,\phi)$ and that of a reference wave $\Psi_\text{{ref}}$:
\begin{equation}
T(\rho,\phi)=|\Phi_{\text{pl}}^{\text{TBB}}(\rho, \phi)+\Psi_\text{{ref}}|^2.
\end{equation} 
In our case, we choose $e^{ik_{x0}x}$, a tilted plane wave, as the reference wave to produce a forked version of the CGEH 
\begin{equation}
T(\rho,\phi)=|N_{pl}J_l(k_{\perp}^{pl}\rho)e^{il\phi}+e^{ik_{x0}x}|^2.
\label{EQN_forkedCGEH}
\end{equation} 
The transverse phase structure of the truncated Bessel beam is shown in Figs. \ref{FIG_TransmissionMasks}(a)-\ref{FIG_TransmissionMasks}(c) for $\Phi_{01}$, $\Phi_{11}$ and $\Phi_{21}$, respectively. Note that the phase structure of the higher order radial modes are divided into $p+1$ annular zones, with the azimuthal phase structure of the $p$-th zone shifted from the central zone by $p\pi$. This, as we will see below, is the key feature to understand of vortex beam generation using high-order diffraction from the CGEH.

On expanding the square in Eq. (\ref{EQN_forkedCGEH}), the transmission function becomes:
\begin{equation}
T(\rho,\phi)=N_{pl}^2J_l^2(k_{\perp}^{pl}\rho)+1+2N_{pl}J_l(k_{\perp}^{pl}\rho)cos(l\phi+k_{x0}x),
\label{EQN_forkedCGEH_expended}
\end{equation} 
which consists of three terms.  The first two terms represent the zeroth order diffraction beam. The last term, in contrast, can be written as the sum of two terms. One is proportional to $e^{i(l\phi+k_{x0}x)}$ and the other to $e^{-i(l\phi+k_{x0}x)}$.  These correspond to the first order diffraction beam and its complex conjugate.  To generate these two diffracted beams experimentally, one needs a greyscale CGEH that can modulate the intensities of the incident plane wave according to $T(\rho,\phi)$, which is technologically challenging.
For ease of practical reproduction, the transmission function is further binarized as 
\begin{equation}
T_{b}(\rho,\phi) = \begin{cases} 1 & \text{for }\; T \ge \alpha T_\text{{max}}\\ 0 & \text{otherwise} \end{cases}
\label{EQN_BinarizedRule}
\end{equation} 
where $T_{b}(\rho,\phi)$ stands for binarized transmission function, $T_\text{{max}}$  is the maximum value of the transmission function given in Eq. (\ref{EQN_forkedCGEH_expended}), and $\alpha$ is a parameter controlling the threshold value used to binarize the gray level representation of the original transmission function and here it is initially set at 0.5. Figures \ref{FIG_TransmissionMasks}(d)-\ref{FIG_TransmissionMasks}(f) show, by way of example, the binarized transmission functions for a CGEH constructed from $\Phi_{01}^{\text{TBB}}$, $\Phi_{11}^{\text{TBB}}$ and $\Phi_{21}^{\text{TBB}}$, respectively. The binarized transmission functions are transferred onto gold-plated silicon nitride membranes using focused ion-beam lithography as shown in Figs. \ref{FIG_TransmissionMasks}(g)-\ref{FIG_TransmissionMasks}(i) for a CGEH constructed from  $\Phi_{01}^{\text{TBB}}$, $\Phi_{11}^{\text{TBB}}$ and $\Phi_{21}^{\text{TBB}}$, respectively.

\begin{figure}
\includegraphics[width=0.75\columnwidth]{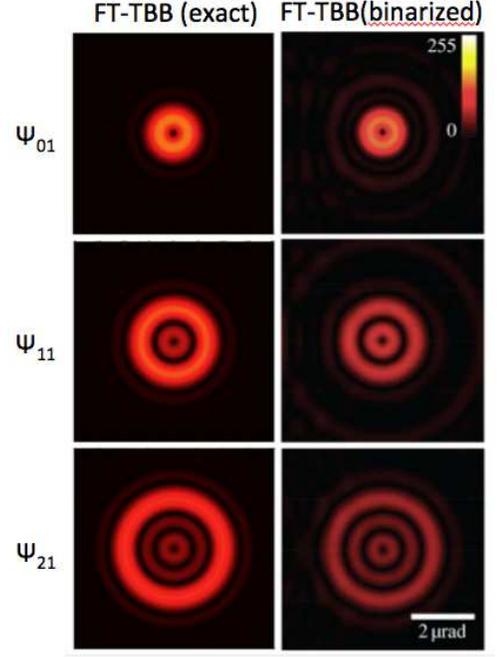}%
\caption{: Shown on the left are the simulated intensity profiles of the FT-TBB evaluated using the exact transmission function as given in Eq. (\ref{EQN_forkedCGEH_expended}).  Shown on the right are the simulated intensity plots of the approximate FT-TBB generated by the binarized version of the transmission function defined by Eq. (\ref{EQN_BinarizedRule}).  The comparison in the top, middle and bottom rows are for the ($p$,$l$) eigenmodes (01), (11) and (21) respectively.}
\label{FIG_comparison_exact_vs_binarization}
\end{figure}

To demonstrate that the binarization process as defined by Eq. (\ref{EQN_BinarizedRule}) does not lead to significant modifications of the FT-TBB beams described by the transmission function as given by Eq. (\ref{EQN_forkedCGEH_expended}), we present a comparison of the results of the simulation of the first-order diffraction pattern with the intensity of the exact wave functions of the FT-TBB beams in Fig. \ref{FIG_comparison_exact_vs_binarization}. The comparison is shown for three typical vortex modes with different $p$ indices, namely, $\Psi_{01}^{\text{FT-TBB}}$, $\Psi_{11}^{\text{FT-TBB}}$, and $\Psi_{21}^{\text{FT-TBB}}$.  The data are presented in the diffraction angle space ($\theta,\varphi$), where $\theta$ is defined as $\arctan{(k_{\perp}/k_0)}$.  The simulation is done for 200-keV electrons and for $\rho_{\text{max}}=2$ $\mu$m. The two results do indeed show that the difference is essentially negligible.  This is consistent with our earlier findings involving the  binary masks of vortex beams with different amplitude functions \cite{Clark2012} ascertaining that in each case the final beam profile is not much changed.  As the main purpose of this paper is about the higher-order diffraction which only exists if a binary transmission function is used, we consider it sufficiently instructive to compare the experimental results with the results of the simulation using the binary transmission function. For brevity, we will henceforth drop the superscript FT-TBB when discussing the beams observed at the focal plane of the apertured electron lens.

\section{Results}
\begin{figure}
\includegraphics[width=1\columnwidth]{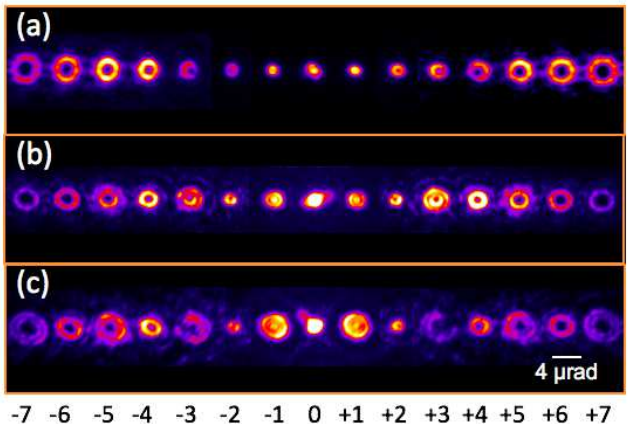}
\caption{Experimental diffraction patterns collected at the back focal plane of the electron microscope, for a CGEH constructed from (a) $\Phi_{01}^{\text{TBB}}$, (b) $\Phi_{11}^{\text{TBB}}$, and (c) $\Phi_{21}^{\text{TBB}}$.}
\label{FIG_FT-TBB}
\end{figure}

Figure \ref{FIG_FT-TBB} shows the electron diffraction patterns recorded at the back focal plane of the binary CGEH generated using the transmission function given in Eq. (\ref{EQN_BinarizedRule}).  The experiment is conducted inside a field-emission transmission electron microscope (JEOL 2200FS).  The grating structures are inserted at the specimen plane and the diffraction patterns are observed at a low-magnification mode to resolve the micronradian diffraction features, with the electron microscope operating at 200 kV in a free lens control mode.  The experimental results are also presented in terms of the diffraction angle whose values are calibrated using the diffraction grating formula $\theta_h \sim \frac{h\lambda}{d}$, where $h$ is the order of diffracting beams, $\lambda$ the electron wavelength, and $d$ the average slit separation.  

\begin{figure*}[htp]
\includegraphics[width=1.75\columnwidth]{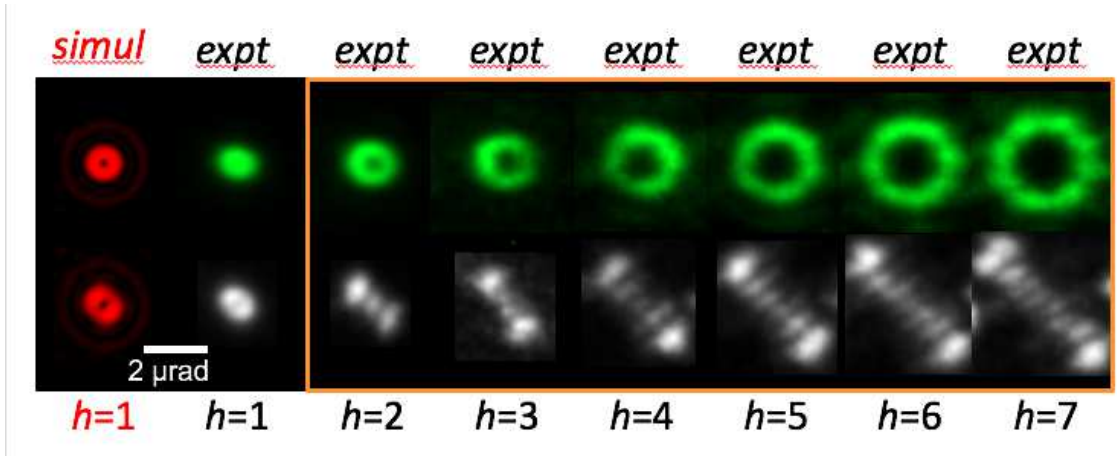}%
\caption{Shown on top are the detailed structures of the diffracted beams and on the bottom their astigmatic transforms for the mask encoding $\Phi_{01}^{\text{TBB}}$.  Here $h$ is the order of diffraction, expt refers to data measured experimentally and simu refers to theoretical simulation.}
\label{FIG_FT-TBB-01}
\end{figure*}

\begin{figure*}
\includegraphics[width=1.75\columnwidth]{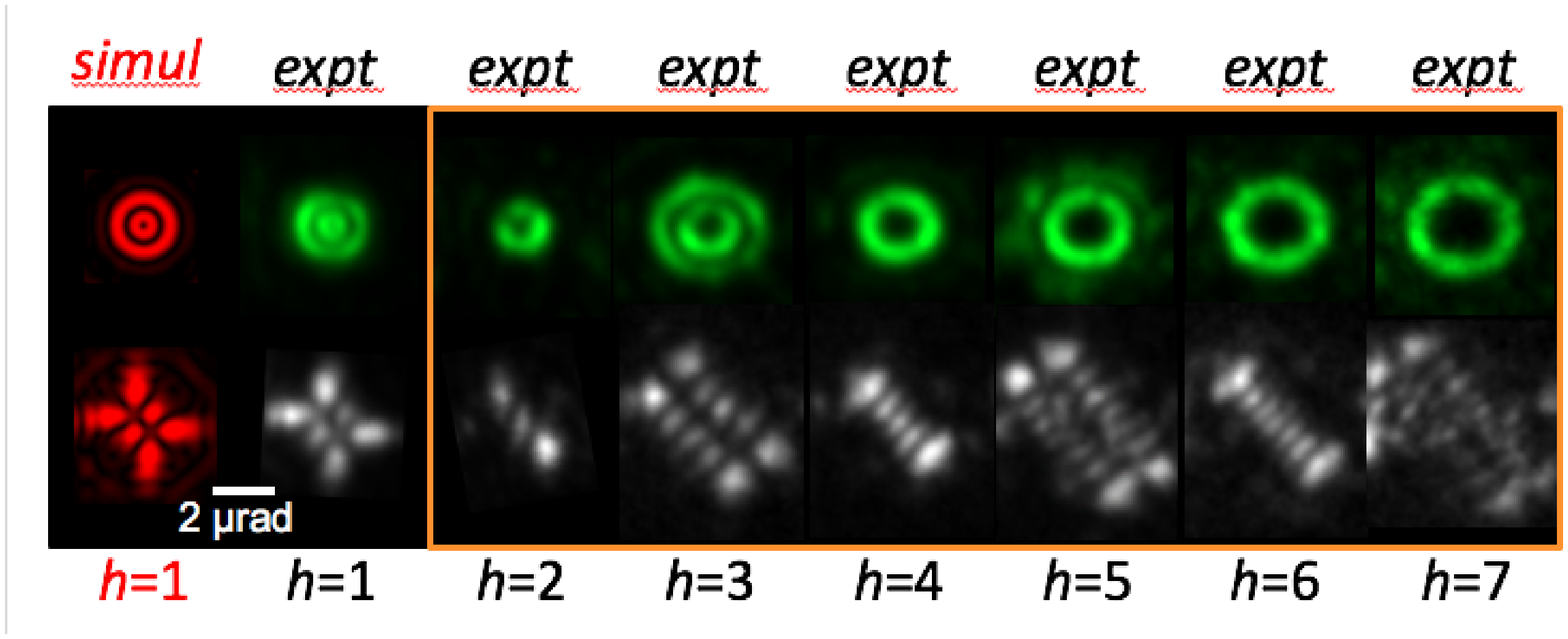}%
\caption{Shown on top are the detailed structures of the diffracted beams and on the bottom their astigmatic transforms for the mask encoding $\Phi_{11}^{\text{TBB}}$.  Here $h$ is the order of diffraction, expt refers to data measured experimentally and simu refers to theoretical simulation.}
\label{FIG_FT-TBB-11}
\end{figure*}

\begin{figure*}
\includegraphics[width=1.75\columnwidth]{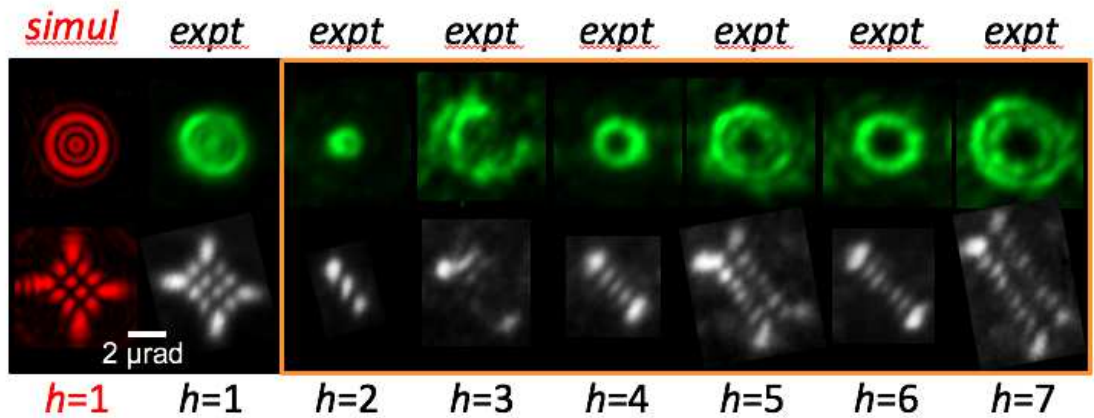}%
\caption{Shown on top are the detailed structures of the diffracted beams and on the bottom their astigmatic transforms for the mask encoding $\Phi_{21}^{\text{TBB}}$.  Here $h$ is the order of diffraction, expt refers to data measured experimentally and simu refers to theoretical simulation.}
\label{FIG_FT-TBB-21}
\end{figure*}

The details of the first-order diffraction patterns have been briefly discussed earlier \cite{Thirunavukkarasu2017} and the main results are summarized here, as shown in the two images on the left side of Figs. \ref{FIG_FT-TBB-01}-\ref{FIG_FT-TBB-21}.  The top image of the second columns in Figs. \ref{FIG_FT-TBB-01}-\ref{FIG_FT-TBB-21} show the respective experimental first-order diffraction (marked by the label $h$=1 and expt).  These compare very well with the corresponding Fourier transforms of the binary transmission functions shown in the top images of the first columns in Figs. \ref{FIG_FT-TBB-01}-\ref{FIG_FT-TBB-21} ($h$=1, simu).  The diffraction pattern of the mask encoding the vortex beam wavefunction $\Phi_{01}^{\text{TBB}}$ has a familiar doughnut shape (Fig. \ref{FIG_FT-TBB-01}).  The corresponding diffraction pattern for the mask, encoding the vortex beam wave function $\Phi_{11}^{\text{TBB}}$, consists of two prominent rings (Fig. \ref{FIG_FT-TBB-11}).  The diffraction pattern of the mask encoding the vortex beam wavefunction $\Phi_{21}^{\text{TBB}}$ shows mainly three concentric rings (Fig. \ref{FIG_FT-TBB-21}), with the outer rings being the brightest. In this way, FT-TBB beams, like TBB beams, are Laguerre-Gaussian-like (LG-like) in the sense that there are always $p$ concentric rings for modes with radial quantum index $p$.  

There are additional weak subsidiary ring structures in all FT-TBB cases.  The results can be understood in terms of the doughnut ring structure multiplied by an Airy-pattern-like point spread function, with the weak subsidiary ring structures corresponding to the sideband structure of the Airy-pattern function. This is because the truncated Bessel beam encoded by the transmission function $T_b$ at the apertured plane can be considered as a product of a proper Bessel beam with its many rings, multiplied by a top-hat function.  The wave at the focal point of the lens is the Fourier transform of the wave at the lens aperture plane.  By the convolution theorem, the Fourier transform of the truncated Bessel beam is the convolution of Fourier transform of the untruncated Bessel beam and the Fourier transform of the top-hat function (which is described by an Airy-pattern function).  One can thus understand the observed (first-order) diffracted beam in terms of an intense ring due to the Fourier transforms of Bessel beams and many subsidiary ring structures due to the convolution of the intense ring structure by the Airy-pattern functions, as shown in Figs. \ref{FIG_FT-TBB-11} and \ref{FIG_FT-TBB-21}.  The inner subsidiary ring or rings are brighter than the outer one due to their smaller radius.  The existence of the weak ring structures distinguishes our bandwidth-limited structured beams from either the Fourier transform of the Bessel beams or the Laguerre-Gaussian beams.

The vortex nature of the phase structure can be demonstrated indirectly using an astigmatic transformation \cite{Beijersbergen1993}.  The astigmatic transformation is achieved inside the electron microscope by introducing astigmatism via astigmatic control \cite{Schattschneider2012b, Shiloh2015}.  The results of the astigmatic transformation are displayed in the bottom panels of Figs. \ref{FIG_FT-TBB-01}-\ref{FIG_FT-TBB-21}.  

The astigmatic probe in the first-order diffraction beam from the mask encoding  $\Phi_{01}^{\text{TBB}}$ consists of two main dots (Fig. \ref{FIG_FT-TBB-01}).  In the case of the higher-order diffraction peaks, a row of increasing numbers of dots is observed and the number of dots increases with the order of the diffracted beam and is such that the number of dots is the number of diffraction order plus one.  Alternatively, the number of dark lines separating the dots equals the number of the diffraction order.  A similar astigmatic transformation is well known in optical vortex beams when LG beams are transformed into Hermite-Gaussian (HG) beams under astigmatic transformation, with the number of dark lines of the Hermite-Gaussian-like electron beam being related to the topological charge of the Laguerre-Gaussian-like electron beam or its intrinsic orbital angular momentum  \cite{Beijersbergen1993, Allen1992}.

For the first-order diffracted beam from the mask encoding  $\Phi_{11}^{\text{TBB}}$ ($\Phi_{21}^{\text{TBB}}$), the astigmatic probe consists of two (three) rows of three (four) dots (Figs. \ref{FIG_FT-TBB-11}( \ref{FIG_FT-TBB-21})).  This is comparable to the optical equivalent of the astigmatic transformation of Laguerre-Gaussian LG$_{\text{pl}}$ beams into Hermite-Gaussian HG$_{\text{nm}}$ beams, where $p=m$ and $l=n-m$ for $n \ge m$.  The number of dark lines separating the bright dots in the HG-like beams can be used to identify the relevant $n$ and $m$ integers characterising the HG$_{\text{nm}}$ beam.  Once identified, they can be used to characterise the corresponding LG$_{\text{pl}}$-like beams.  In our experimental results (Figs. \ref{FIG_FT-TBB-11} and \ref{FIG_FT-TBB-21}), we observed the intensity patterns of HG$_{21}$-like (HG$_{31}$-like) beams after astigmatic transformation, as there are 2$\times$1 (3$\times$2) dark lines separating the bright dot-like intensity patterns. This is consistent with the original LG$_{11}$-like (LG$_{21}$-like) beams we started with.  It is important to emphasise that the diffracted beams we have observed are not exactly the same as LG beams as the actual intensity patterns here have additional weaker features that can be seen when the intensity of the images is increased.

Applying the analysis outlined above on the intensity patterns of the astigmatically transformed high-order diffracted beams shown in Fig. \ref{FIG_FT-TBB-01}, we find that the emerging HG-like beams have dark lines separating the single row of bright dots in proportion to the order of the diffraction.  This suggests that the radial indices of the untransformed diffracted beams remain as $p=0$ while the angular index scales with the order of the diffracted beam.  In other words, the $h$th order diffraction of the original mask encoding  $\Phi_{0l}^{\text{TBB}}$ can be classified as an LG$_\text{{0(hl)}}$-like vortex beam. This effect has been exploited to produce electron vortex beams with very high topological charge \cite{McMorran2011}. 

Can we use the high-order diffraction from our CGEH gratings to produce electron beams with higher radial indices as well?  Our results demonstrate that this is not possible.  In Fig. \ref{FIG_FT-TBB-11}, the diffracted beams with $h$ = 2, 4 and 6 can be similarly identified by their astigmatically transformed beam pattern as LG$_{\text{0h}}$-like vortex beams.  The diffracted beams with $h$ = 3, 5 and 7 can be roughly identified with LG$_{\text{1h}}$-like vortex beams.  This suggests that the higher-order diffraction still generates vortex beams with $h$-times the topological charge of the original beam encoded in the CGEH, however the radial indices either stay the same (for odd $h$) or are effectively reduced to $p=0$ for even $h$.  A very similar effect is observed in Fig. \ref{FIG_FT-TBB-21}.  For even $h$, the radial indices of the corresponding beams are both zero, despite the fact the radial index of the vortex beam encoding the CGEH is 2.  The experimental result for $h=3$ is not perfect, but it still gives an indication that it is closer to that found in the first-order diffracted beam.  Together with other high odd-order diffracting beams, their radial indices are about the same as the first order diffracting beams. 
\section{Discussion}
A key observation is that the angular indices of the higher-order diffracted beams scale with the diffraction order $h$ but the radial indices of the diffracted beams revert to zero for even $h$ and remain unchanged for odd $h$.  This can also be understood in terms of phase amplification seen in higher-order diffraction.  The phase amplification effect is a well-known phenomenon in holography \cite{Tonomura2013} and can be understood from the basic principles of the diffraction hologram.  In our case, the phase amplification stems from the binarization of the transmission function shown in Eq. (\ref{EQN_forkedCGEH_expended}).  According to the binarization rule given in Eq. (\ref{EQN_BinarizedRule}), the binarized transmission function consists of a rectangular grating structure as shown in Fig. \ref{FIG_TransmissionMasks}(d)-\ref{FIG_TransmissionMasks}(f).  In analogy with the mathematics for the regular rectangular holographic grating (see the Appendix), the transmission function for a simple vortex beam of the form $e^{il\phi}$ can be written as
\begin{equation}
T_l=|e^{-il\phi}+e^{ik_{x0}x}|^2=2[1+\cos(k_{x0}x+l\phi)]
\end{equation}
and the corresponding binary transmission function is a series of slits with width $d-a$ and repetition length $d$ (which is related to the wavelength of the reference wave),
\begin{equation}
T_{lb}=\frac{a}{d}+\sum_{h=1,2,...} \frac{2}{h\pi}\sin(\frac{h\pi a}{d})\cos(hk_{x0}x+hl\phi).
\end{equation}
The resulting diffraction pattern is given by
\begin{widetext}
\begin{equation}
\Psi(k_x, k_y)={\cal{F}}(T)=\frac{a}{d}\delta(k_x, k_y)+ \sum_{h=\pm 1, \pm 2, ...}\frac{2}{h\pi}\sin(\frac{h\pi a}{d})\delta(k_x-hk_{x0}, k_y)\circledast {\cal{F}}(e^{ihl\phi}).
\end{equation}
\end{widetext}
where ${\cal{F}}$ is the Fourier operator and $\delta(k_x, k_y)$ is shorthand for $\delta(k_x)\delta(k_y)$.
We have also used the convolution theorem to write
\begin{equation}
{\cal{F}}(e^{ihk_{xo}x+il\phi})={\cal{F}}(e^{ik_{xo}x} e^{il\phi})\propto\delta(k_x-hk_{x0},k_y)\circledast {\cal F}(e^{il\phi}).
\end{equation}
It is known that, due to rotational invariance \cite{Wang2009d}, the Fourier transform of the vortex beam phase function is another beam with a similar vortex phase function form \cite{Schattschneider2012e}:
\begin{equation}
{\cal F}(e^{il\phi}) \propto e^{il\varphi}.
\end{equation}
where $\varphi$ is the azimuthal variable in the cylindrical polar coordinate description of the diffraction plane $(k_{\perp}, \varphi)$ or alternatively $(k_x, k_y)$.

\begin{figure}
\includegraphics[width=0.7\columnwidth]{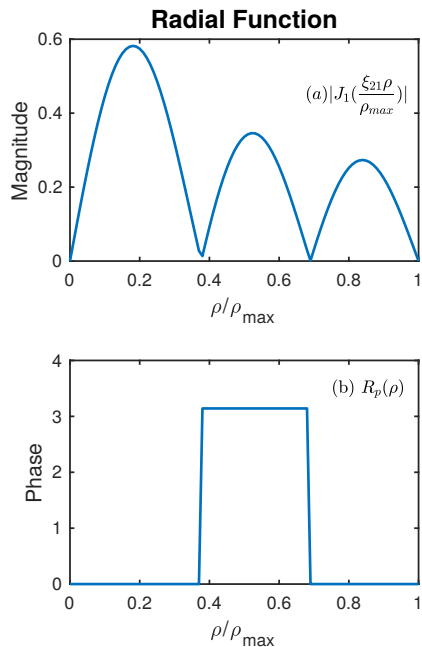}%
\caption{Radial dependence of the wave function for the mode  expressed in terms of (a) the amplitude $\lvert J_1(\frac{\xi_{21}\rho}{\rho_{\text{max}}})\rvert$ and (b) the phase $R_p(\rho)$ in the radial direction.}
\label{FIG_PhaseLineProfile}
\end{figure}

For a diffractive hologram encoded with a vortex beam with an azimuthal phase ramp term $e^{il\phi}$, the same phase structure is reproduced in the first-order diffraction as $e^{il\varphi}$; the phase amplification effect of the $h$th-order diffracted beam will result in a vortex beam in the diffraction plane with an azimuthal term $e^{ihl\varphi}$. This is the mathematical basis for the generation of electron vortex beams with large topological charges using phase amplification associated with high-order diffraction.  
The situation for the radial indices of the reconstructed beam is rather different and can be understood by an examination of the radial variation as shown in Fig. 1.  The higher-order modes are different from the $p=0$ mode by the division of the apertured area into concentric zones; each differs from its immediate neighbour by a negative sign.  Alternatively, this can be considered as a phase change of $\pi$.  This can be understood if we write the radial dependence of the TBB as
\begin{equation}
J_l(k_{\perp}^{pl}\rho)=|J_l(k_{\perp}^{pl}\rho)|e^{iR_p(\rho)}.
\end{equation}
An example of the phase function for the radial function $R_p(\rho)$ for the mode $\Psi_{21}$ is shown in Fig. \ref{FIG_PhaseLineProfile}.
The effect of this phase change is also evident in the amplitude mask function as the fringe patterns in the neighbouring zones are shifted to be complementary to each other (see Fig. \ref{FIG_TransmissionMasks}(d)-\ref{FIG_TransmissionMasks}(f).  This means that the binary transmission function $T_b(p,l)$ for the two-dimensional orthonormal modes with non-zero radial indices can be written as 
\begin{equation}
T_b(p,l)=\frac{a}{d}+\sum_{h=1} \frac{2}{h\pi}\sin(\frac{h\pi a}{d})\cos[hk_{x0}x+hl\phi+hR_p(\rho)]
\end{equation}
such that
\begin{widetext}
\begin{equation}
\Psi_{pl}(k_x, k_y)={\cal{F}}(T_b(p,l))=\frac{a}{d}\delta(0,0)+ \sum_{h=\pm 1, \pm 2, ...}\frac{2}{h\pi}\sin(\frac{h\pi a}{d})\delta(k_x-hk_{x0}, k_y)\circledast {\cal{F}} (e^{ihl\phi})\circledast {\cal{F}} (e^{ihR(\rho)}).
\label{Eq_binary_FT_TBB}
\end{equation}
\end{widetext}

The even-odd effect seen clearly in the higher-order diffracting beams, particularly in their astigmatic transformed states, can now be explained in terms of the effect of phase amplification on the last Fourier-transformed factor in Eq. (\ref{Eq_binary_FT_TBB}).  In the case of the even ($h=2k$)-order diffracted beams, the phase change of $\pi$ between the neighbouring annular zones in the grating mask will become a phase change of $2k\pi$, where $k$ is a positive integer.  This effectively results in the elimination of the sign-alternating phase zone structure, because $e^{i2kR(\rho)}=1$, and hence this results in the observation of the beam mode with the ($p=0$)-like radial structure.   In the case of the odd  ($h=2k+1$)-order diffraction beams,  $e^{i(2k+1)R(\rho)}=e^{iR(\rho)}$, the phase amplification results in no change of the phase of the zone structure, so the radial indices of the resulting higher-order diffracted beam will be the same as that of the first-order one.

A closer examination of the fine structures of the higher-order diffraction patterns suggests an occasional departure from the simple even-odd rule.  This can be partially understood in terms of missing information in the phase amplification process, as the zeros of the Bessel functions represent the missing information at their radial positions in the diffraction plane.  This means that the phase amplification by an even number of times cannot strictly transform the radial wave function back to the zeroth-order wave function.  Also, the lens aberration, whose amplitude also scales with the radial variable of the beam, will also modify the radial dependence of the beam.  In other words, a beam generated by an even-order diffraction is no longer in a pure state, which may explain the complex astigmatic pattern for $h=6$, as seen in Fig. \ref{FIG_FT-TBB-21}.  However, judging by the experimental astigmatically transformed results, the departure from a pure state for the cases we have examined in this paper is generally quite small.

In addition, the overall intensities of the diffracted beams also depend on the details of the binarization, which have not been explicitly mentioned in the above discussion of the even-odd rule. We see that the intensity profile of the third-order diffraction shown in Fig. \ref{FIG_FT-TBB-21} does not follow the above simple even-odd rule.  What we have not mentioned explicitly so far is the envelope effect of the beam-blocking bars in the binary CGEH, the width of which is defined by the  parameter $\alpha$ defined in the binarization rule given in Eq. (\ref{EQN_BinarizedRule}).  The shape functions of these bars will produce their own Fourier transforms which affect the intensities of the resulting diffracted beams through its envelope function $\sin(\frac{h\pi a}{d})$. In our case, $a/d\sim 0.4$ as seen in Fig. \ref{FIG_TransmissionMasks}, which suggests that the envelope function reaches a minimum near $h=3$.  The overall effect on the resulting envelope function can be clearly seen in Fig. \ref{FIG_FT-TBB} as a overall depression of the intensity of the third-order diffracted beam.  Due to the different phase shifts, the shape function of the binary CGEH varies at different zones in the mask, resulting in different effects on the intensities at different radial parts of the diffracted beams.

One application of our two-dimensional cylindrical basis set is in the efficient description of arbitrarily shaped electron probes.  This is, for example, necessary in electron ptychography \cite{Cao2016}, scanning transmission electron microscopy \cite{Crewe1970}, and recently developed imaging scanning transmission electron microscopy \cite{Rosenauer2014}.  Our biorthogonal beam modes constitute a special example of a set of structured beams showing superoscillating electron wave functions with sub-diffraction features \cite{Remez2017}.   Such structured electron probes, particularly those represented by the topologically non-trivial $l \neq 0$ and $p \neq 0$ beams, have not been explored much in scanning probe microscopy and spectroscopy, but may hold advantages over their more familiar topologically trivial counterparts. For example, their distributed beam intensities may minimize the beam-induced atomic motion during single-atom imaging and spectroscopy \cite{Senga2015, Ramasse2013}.
\section{Conclusions}
In summary, we have presented the procedure and results of an experimental study aimed at the production of cylindrical orthonormal electron modes using binary computer generated holograms.  In particular, we found that higher-order diffraction beams can produce higher-order azimuthal electron-beam modes from binary masks encoding a lower-order azimuthal electron-beam mode through a phase amplification effect.  This useful property also applies to the beams with high-order radial indices.  The same phase amplification effect however leads to a complex variation in the higher-order radial modes of the electron beams.  All the even-order diffracted beams produce approximate beams with a zeroth radial index, while the radial structure of the original beam should be largely reproduced for the odd-order diffracted beams.  Additional factors such as the binarization threshold and the shape of the fringes also contribute to the final shape.  Furthermore, the overlap of different diffracted beams for very large diffraction orders is an additional factor to consider in the interpretation of the experimental results.  This suggests that the high-order diffraction from the CGEH involving a discrete $\pi$-step change in the radial phase structure is not a viable way to generate beams with higher-order radial structures, in contrast to the azimuthal modes.  It is important to point out that this conclusion is quite general; for example, it should also apply to the holographic reproduction of Laguerre-Gaussian modes \cite{Allen1992} often encountered in optical vortex beams. The same non-trivial effect will lead to complex changes to the radial structure of the beams arising from the superposition of the orthonormal set of modes we have discussed here.

\section{Acknowledgement}

This research was supported by the UK Engineering and Physical Sciences Research Council Grant (No.EP/J022098/1).

\section*{Appendix}
\renewcommand{\theequation}{A-\arabic{equation}}  
\setcounter{equation}{0} 
By a regular rectangular holographic grating we mean a grating derived by the binarization of a transmission function due to the interference of a plane wave (the encoding wave) with a tilted plane wave (the reference wave).  In analogy to Eq. (\ref{EQN_forkedCGEH}), the corresponding transmission function is given by:
\begin{equation}
T=|1+e^{ik_xx}|^2=2[1+cos(k_{x0}x)]
\label{EQN_ATF}
\end{equation}
where $k_{x0}=2\pi/d$ and $d$ is the wavelength of the reference wave.  Applying the binarized rule of Eq.(\ref{EQN_BinarizedRule}), the binarized transmission function consists of regularly spaced rectangular gaps of width $a$, at a repeat distance $d$ apart.  The width $a$ is a function of the parameter $\alpha$ defined in Eq. (\ref{EQN_BinarizedRule}).  At $\alpha=0.5$, we have $a=0.5d$ and the grating structure has a square-wave profile along the $x$-direction.  Otherwise, the amplitude mask has a rectangular line profile along the $x$-direction. 

Expanding the periodic and rectangular wave in terms of the cosine series  we have the binarized version of the transmission function $T_b$, represented by
\begin{equation}
T_b=\frac{a}{d}+\sum_{h=1} \frac{2}{h\pi}\sin(\frac{h\pi a}{d})\cos(hk_{x0}x)
\label{EQN_BTF}
\end{equation}
The diffraction of the incident beam by the mask defined by a transmission function $T$ is the Fourier transform of $T$.  It helps to recall that
\begin{equation}
\cos(k_{x0}x)=\frac{1}{2} (e^{ik_{x0}x}+e^{-ik_{x0}x})
\end{equation}
and
\begin{equation}
{\cal{F}}(e^{\pm ik_{x0}x})\propto \delta(k_x\mp k_{x0})
\end{equation}
For the analogous transmission function defined in Eq. (\ref{EQN_ATF}), the wave function at the diffraction plane is given by:
\begin{widetext}
\begin{equation}
\Psi(k_x, k_y)={\cal{F}}(T)\propto 4\delta(k_x)\delta(k_y)+2\delta(k_x-k_{x0})\delta(k_y)+2\delta(k_x+k_{x0})\delta(k_y)
\end{equation}
\end{widetext}
The first term corresponds to the non-diffraction beam, while the last two terms consist of the diffracted beam and its complex conjugate.  For simplicity, we redefine $\delta(k_x)\delta(k_y)$ as $\delta(k_x, k_y)$.  Similarly, we have $\delta(k_x\pm k_{x0})\delta(k_y)$ as $\delta(k_x\pm k_{x0}, k_y)$.

For the binarized transmission function, the corresponding wave function in the diffraction plane is given by:
\begin{widetext}
\begin{equation}
\Psi(k_x, k_y)={\cal{F}}(T)\propto \frac{a}{d}\delta(k_x,k_y)+ \sum_{h=\pm 1, \pm 2, ...}\frac{2}{h\pi}\sin(\frac{h\pi a}{d})\delta(k_x-hk_{x0}, k_y)
\end{equation}
\end{widetext}
The binarization processes are responsible for the higher-order diffracted beams at $\delta(k_x-hk_{x0}, k_y)$ for $|h|>1$.  For the square-shaped regular diffraction pattern, $a=0.5d$ and $\sin(\frac{h\pi a}{d})=0$ for even values of $h$, hence only terms with odd values of $h$ are involved. This means that the intensities of the even-order diffraction beams are zero or very small.  

In our non-trivial case, because the hologram consists of interference patterns with locally variable bar width, the even-order beams are generally not completely eliminated but are related to the even-odd effect we observe in the change to the radial indices.

%

\end{document}